\documentclass[a4paper,11pt]{article}
\pdfoutput=1 

\usepackage{jheppub} 

\usepackage[T1]{fontenc} 
\usepackage{grffile}
\usepackage{braket}
\usepackage{subcaption}
\usepackage[compat=1.0.0]{tikz-feynman}
\title{Quantum vacuum effects on the formation of black holes}

\author[a]{Moslem Shafiee}
\author[a,b]{and Yousef Bahrampour}

\affiliation[a]{Faculty of Mathematics and Computer, Shahid Bahonar University of Kerman,\\
P.O. Box 76169-13639, Kerman, Iran}
\affiliation[b]{Mahani Mathematical Research Center,\\
P.O. Box 76188-68366, Kerman, Iran}

\emailAdd{m.shafiee@math.uk.ac.ir}
\emailAdd{bahram@uk.ac.ir}

\abstract{We study the backreaction of quantum fields induced through the vacuum polarization and the conformal anomaly on the collapse of a thin shell of dust. It is shown that the final fate of the collapse process depends on the physical properties of the shell, including its rest and gravitational masses. Investigating the conditions for the formation of black holes, we notice that quantum effects modify the geometry and structure of Schwarzschild space-time  in such a way that black holes have two horizons, an inner and an outer horizon. If the gravitational mass of the shell is about that of an ordinary star, then in most cases, the semi-classical collapse will terminate in a singularity, and in general, quantum fluctuations are not strong enough to prevent the creation of the singularity. Although under certain conditions, it is possible to form a non-singular black hole, i.e., a regular black hole. In this way, the collapse stops at a radius much larger than the Planck length below the inner horizon, and the shell bounces and starts an expansion.}

\begin{document} 
\maketitle
\flushbottom
\raggedbottom
\section{Introduction}

Since the emergence of the theory of general relativity and the prediction of the existence of black holes, the question that how and under what conditions these structures are formed in the real world has always been raised. In one of the most basic and plain arguments, Oppenheimer and Snyder indicated that a pressureless ball of perfect fluid can shrink under its own gravity so that all of its components eventually collapse into a single point of infinite density, i.e., a singularity \cite{a}. Later research by Penrose showed that in the general case and without considering spherical symmetry or any specific type of matter, when an object is compressed to a certain extent (to its Schwarzschild radius), then all of its ingredients will gather at a single point, terminating in a singular state, and ultimately that object will become a black hole \cite{b}. Also, the resulting singularity is always  concealed behind an event horizon from a distant observer. In other words, general relativity predicts that the final outcome of the gravitational collapse of a sufficiently massive body is a singular point that is shielded by an event horizon.

But by taking into account the quantum theory, it seems that this classical view of gravitational collapse is not valid, and the classical scenario undergoes fundamental modifications. Quantum effects may cause gravitational collapse to stop before the creation of a singularity, leading to the production of non-singular black holes. This issue has been discussed and investigated a lot in recent decades, however, a clear and complete picture has not emerged yet. In the first proposed model for a non-singular black hole, Bardeen found a spherically symmetric solution to Einstein's equations coupled to a nonlinear electrodynamic \cite{c}. In this model, the central singularity is removed and replaced by a core of the charged matter, and there may be two horizons instead of one. This scenario has been researched in recent decades and has been extended to uncharged and rotating non-singular black holes, which are available in \cite{d1, d2, d3, d4, d5, d6, d7, d8, d9, d10, d11, d12, d13}. A detailed review of these works and further references on the subject can be found in \cite{d14, d15, d16, d17}.

A general mathematical approach and description of regular, static, spherically symmetric, and asymptotically flat space-times for characterizing non-singular models of black holes has been initially introduced and explored by Hayward \cite{e1}. The metric for such space-time has a finite curvature at the center and is de Sitter-like around this area, and it converts to Schwarzschild geometry in far distances. There is also a trapped region with two boundaries, an outer horizon located near the Schwarzschild radius, and an inner horizon which is determined with a parameter relating to a cosmological constant. The Hayward model has been generalized to establish a compatible framework for the evaporation of non-singular black holes \cite{d15, d16}.

There is an argument that the expectation values of the stress-energy tensor for quantum fields in curved space-times and in presence of event horizons are in the Hadamard form \cite{e0, e}. This means that the divergencies in the stress-energy tensor can be removed by ordinary methods of renormalization, so its components have finite values. Hence at the semi-classical level, the endpoint of gravitational collapse may not be different from the classical picture. Of course, this does not clearly answer the question of whether quantum effects allow the formation of an event horizon. In addition, it has been discussed that even if the quantum state is Hadamard, if the collapse does not happen in a free-fall way (i.e., be accelerated), then it permits the renormalized stress-energy tensor to grow up \cite{e3}. Indeed, the dynamics of space-time around a collapsing object causes quantum effects such as vacuum polarization to intensify \cite{e2}. These effects delay the collapse process and may stop it even before the horizon is formed. In this case, the amount of redshift from the surface of the created object (a black star) is so high that it looks like an event horizon. Also, the geometry around a collapsing body gives rise to Hawking-like radiation, which can prevent the formation of an event horizon \cite{e4, e5, e6}. In this way, the corresponding issues and problems such as the information loss paradox that arises in the presence of horizons will no longer be proposed. Although in this regard, there are also criticisms that this Hawking-like radiation is negligible and is not capable of altering the final outcome of a gravitational collapse \cite{e7, e8}.

The issue of the formation of singularities and the evaporation of black holes has been also discussed and investigated based on the framework of a quantum gravitational approach \cite{fb, f2, f, f3, f1}. The Heisenberg uncertainty principle acts as a repulsive force against the compression of matter. With more compression of matter, this force also becomes bigger until it finally overcomes gravity. Therefore, in the final stages of a star's gravitational collapse, the quantum gravitational effects will hinder the process to carry on \cite{f4}. In this case, when the density of matter reaches the Planck density, the collapse stops and enters an expansion phase from the contraction (a bounce). The bounce radius can be much larger than the Planck radius because the effects of quantum gravity turn on when they reach the level of the Planck energy, not the Planck length. The phase change occurs in the frame of the star itself in a short and finite time, while it would be very long for an external observer due to the large time dilation, and thus from the outside, it would appear as a stationary black hole (a Planck star). In the final stages of Hawking radiation and evaporation of this star, a remnant will remain, so the information will not be destroyed. These phenomena emit signals through the space that can be detectable and observable \cite{f5, f6}. A string view on the subject is also proposed in \cite{f7}.

In this paper, the problem of gravitational collapse for a spherical shell consisting of pressureless dust is investigated by considering quantum effects. In section \ref{section 1}, the expectation value of the stress-energy tensor of quantum fields in the vacuum state is introduced in the Schwarzschild background. The calculations require specifying a vacuum state for quantum fields. The true vacuum state for a collapsing body is a state in which particles were absent in the past null infinity. The Unruh vacuum has this feature, so the expectation value of the stress-energy tensor is considered in this state instead of other states such as the Hawking-Hartle vacuum. The stress-energy tensor can be divided into three subparts, each of which represents a quantum effect. The part related to its trace is determined by the conformal anomaly, which depends on the curvature of space-time and the properties of quantum fields, specifically their spin. The other components of the stress-energy tensor are associated with two quantum effects: 1-Hawking radiation and 2- vacuum polarization. For the scenario to be close to reality, the mass of the shell is considered to be about that of an ordinary star, the mass of the sun. Since the Hawking radiation is very small for such objects \cite{Hawking1, Hawking2}, we have neglected it in the calculations. Further, it has been shown that Hawking radiation is not strong enough to overcome the gravitational collapse alone, but only slows down and delays the process \cite{Paranjape}.

The calculation of vacuum polarization in the Schwarzschild geometry indicates that its value is highly dependent on the distance from the gravitational source. As the radius of the shell decreases, the curvature of space-time increases in its vicinity, so the vacuum polarization also grows, and as a result, near the Schwarzschild radius of the shell, the components of the stress-energy tensor will have significant values, which can be decisive in the final fate of the collapse. For the semi-classical method to be valid, all components of the stress-energy tensor are calculated only in the first order of $\hbar$ because it has been argued that due to the nature of quantum theory, higher orders are not reliable and can lead to incorrect results \cite{Nomura}.

Next, by solving Einstein's equations in the presence of the obtained stress-energy tensor, we find the modified Schwarzschild solution. In this case, some terms proportional to $1/r^2$, $1/r^3$, and $1/r^4$ are added to the Schwarzschild metric, resulting in a trapped surface with two outer and inner horizons as boundaries. The outer horizon is located at a radius close to the classical Schwarzschild radius (slightly smaller) and the inner horizon is a multiple of $l_p$ with the factor of $(\dfrac{M}{m_p})^\frac{1}{3}$, which is much larger than the Planck length for astrophysical black holes.

In section \ref{section 3}, we study the gravitational collapse of a spherical shell by examining its internal and external geometry that have been altered by quantum effects. By determining the equations governing the dynamics of the shell, we will see that the ratio of its rest mass to its gravitational mass has a crucial role in the final outcome of the collapse. If this ratio is around $2$, then the gravitational collapse ceases near the position of the outer horizon, and therefore neither a horizon nor a singularity is formed. In the case where the rest mass of the shell is zero or the value of the ratio is very close to it, the collapse at a distance below the inner horizon becomes an expansion state, in other words, a bounce occurs. Therefore, the gravitational singularity will no longer create, and the final product will be a non-singular black hole. In general, when the ratio has a value other than the mentioned cases, a black hole with a central singularity is formed. This means that the semi-classical method predicts that the final result of the gravitational collapse of a sufficiently massive star will be a black hole. Although in this picture, there are differences with the classical scenario, including the existence of two horizons rather than one horizon. Also, due to the backreaction of the quantum fields, the process of gravitational collapse in this case will be longer than the classical view and proceeds with delay.

In all calculations, we use the units in which $G=c=\hbar=m^2_p=l^2_p=1$. Of course, to show the order of quantum effects, we have put $m_p$ in all relations and $l_p$ in some cases.
\section{Quantum-corrected Schwarzschid geometry}\label{section 1}
Fluctuations in quantum fields, originated from the uncertainty principle, require the vacuum state to have a non-zero energy value. This issue causes Einstein's equations to be rewritten in the absence of matter as 
\begin{equation}\label{1}
G_{\mu\nu}=8\pi\braket {T_{\mu\nu}},
\end{equation}
where $\braket {T_{\mu\nu}}$ is the expectation value of the stress-energy tensor for quantum fields. Thus, one expects that by considering the quantum effects, some modifications in the solutions of Einstein's equations will appear. Here we aim to find the spherically symmetric solution of \eqref{1}, i.e., the quantum-corrected Schwarzschild solution. In the classical case, this solution is described through the metric
\begin{equation}\label{cl metric}
ds^2=-(1-\frac{2M}{r})dt^2+(1-\frac{2M}{r})^{-1}dr^2+r^2(d\theta^2+\sin^2\theta d\phi^2).
\end{equation}

By applying the spherical symmetry, the stress-energy tensor will have four independent non-zero components: an energy density $\rho=-\braket{T^t_t}$, an energy flux density $l=-\braket{T^r_t}/(1-2M/r)$, a radial stress $p_\Vert=\braket{T^r_r}$, and a transverse stress $p_\bot=\braket{T^\theta_\theta}=\braket{T^\phi_\phi}$. These components are also related to each other through the local energy-momentum conservation equation $\nabla_\nu\braket {T^{\mu\nu}}=0$. The stress-energy tensor consisits of two subparts \cite{g1, g2, g3, g4}: a trace determined by conformal anomaly, and a traceless regular part. Since these two subparts have different physical origins, they satisfy the local energy-momentum conservation condition seprately, i.e., $\nabla_\nu\braket {T^{\mu\nu (ca)}}=\nabla_\nu\braket {T^{\mu\nu (reg)}}=0$. The regular subpart itself contains two quantum effects: a traceless diagonal part representing the vacuum polarization, and a non-diagonal part describing the Hawking radiation.

The conformal anomaly has been calculated in different methods for various quantum fields \cite{s11, s12, s13, s14}. Because massive particles have a much smaller contribution than massless particles in the expectation value of the stress-energy tensor for the vacuum state, so in this regard, their influence can be ignored. The general form of the conformal anomaly for all type of massless fields is \cite{s15}
\begin{equation}\label{2}
\braket{T^\mu_\mu}=\frac{m_p^2}{2880\pi^2}\sum q_s\big(C_{\mu\nu\rho\sigma}C^{\mu\nu\rho\sigma}+R_{\mu\nu}R^{\mu\nu}-\frac{1}{3}R^2\big),
\end{equation}
where $C_{\mu\nu\rho\sigma}$, $R_{\mu\nu}$, and $R$ are the Weyl tensor, the Ricci tensor, and the Ricci scalar, respectively. $q_s$ is a constant number whose value depends on the spin of the quantum field, which is equal to $1$, $-13$, and $212$ for $s=0, 1, 2$, respectively. In Schwarzschild geometry, we have $C_{\mu\nu\rho\sigma}C^{\mu\nu\rho\sigma}=48M^2/r^6$ and $R_{\mu\nu}=R=0$. Therefore, in the first order of $\hbar$ the relation of \eqref{2} takes the form
\begin{equation}\label{4}
\braket {T^\mu_\mu }=\frac{m^2_p}{60\pi^2}\frac{M^2}{r^6}\sum q_s.
\end{equation}

The conformal anomaly respect to all the symmetries of the Schwarzschild space-time \cite{g2, g3, g4}. The invariance of the Schwarzschild curvature tensor under radial boosts implies that $\rho^{ca}=-p^{ca}_\Vert$. The spherical symmetry also gives $\braket{T^{\theta (ca)}_\theta}=\braket{T^{\phi (ca)}_\phi}=p^{ca}_\bot$. Inserting these conditions into the local energy-momentum conservation relation, one finds
\begin{equation}\label{5}
p^{ca}_\Vert-p^{ca}_\bot+\frac{r}{2}\partial_rp^{ca}_\Vert=0.
\end{equation}

Using \eqref{4} and \eqref{5}, the components of the conforaml anomalous subpart are specified 
\begin{subequations}\label{ca}
\begin{equation}
\rho^{ca}=-p^{ca}_\Vert=\frac{m^2_p}{120\pi^2}\frac{M^2}{r^6}\sum q_s,
\end{equation}
\begin{equation}
p^{ca}_\bot=\frac{m^2_p}{60\pi^2}\frac{M^2}{r^6}\sum q_s.
\end{equation}
\end{subequations}

Other components of the stress-energy tensor are associated with the Hawking radiation and the vacuum polarization. The contribution of Hawking radiation for a body with the mass of the sun is negligible, therefore we do not include it in the calculations (see \cite{g3} and \cite{g4}). The vacuum polarization has been studied in the Schwarzschild background of \eqref{cl metric} for different quantum vacuum states, includinng the Unruh vacuum \cite{vp1, vp2, vp3, vp4}. It is shown that the components of the vacuum polarization subpart of the stress-energy tensor in the first order are \cite{g2, vp5, vp6}
\begin{equation}\label{vp}
\begin{split}
\rho^{vp}=-p^{vp}_\Vert=p^{vp}_\bot=&\frac{m^2_p}{7680\pi^2}\frac{D}{r^4}\sum A _s+\frac{m^2_p}{1280\pi^2}\frac{M}{r^5}\sum A_s(B_s-1)\\
&-\frac{m^2_p}{768\pi^2}\frac{M^2}{r^6}\sum A_sB_s.
\end{split}
\end{equation}
$D$ is an integration constant with the value of $0.621$. Coefficients of $A_s$ and $B_s$ are determined by particle's spin, which have values of $(14.26, 6.49, 0.74)$ and $(0.54, 3.8, 25)$, respectively, for $s=0, 1, 2$. Finally, using relations of \eqref{ca} and \eqref{vp}, it is possible to determine the desired components of the stress-energy tensor 
\begin{equation}
\begin{split}
\braket{T^t_t}=-\rho&=-(\rho^{ca}+\rho^{vp}),\quad \braket{T^r_r}=p_\Vert=p^{ca}_\Vert+p^{vp}_\Vert,\\
&\braket{T^\theta_\theta}=\braket{T^\phi_\phi}=p_\bot=p^{ca}_\bot+p^{vp}_\bot.
\end{split}
\end{equation}

Now we find the spherically symmetric solutions to equation of \eqref{1} for the above stress-energy tensor. We assume the general form of the metric for the sphericalliy symmetric space-time as
\begin{equation}\label{gmetric}
ds^2=-f(r)dt^2+g(r)dr^2+r^2(d\theta^2+\sin^2\theta d\phi^2),
\end{equation}
where $f$ and $g$ are functions of r, which must be specified \footnote{If we also suppose $f$ and $g$ as functions of $t$, then Einstein's equation $G^r_t=0$ concludes that these two functions are independent of $t$. So from the beginning, they are considered only functions of $r$.}. Inserting the metric of \eqref{gmetric} in \eqref{1} yields
\begin{subequations}\label{e equation}
\begin{equation}\label{first}
G^t_t=\frac{1}{r^2g}-\frac{g^\prime}{r^2g}-\frac{1}{r^2}=-8\pi\rho,
\end{equation}
\begin{equation}\label{second}
G^r_r=\frac{1}{r^2g}+\frac{f^\prime}{rfg}-\frac{1}{r^2}=8\pi p_\Vert,
\end{equation}
\begin{equation}\label{third}
G^\theta_\theta=G^\phi_\phi=\frac{1}{4rf^2g^2}\big(2fg(rf^\prime)^\prime-rf^\prime(fg)^\prime-2f^2g^\prime\big)=8\pi p_\bot,
\end{equation}
\end{subequations}
where prime denotes the derivative with respect to $r$. From the first equation, one finds
\begin{equation}
(\frac{r}{g})^\prime=1-8\pi r^2\rho.
\end{equation}
Therefore, we get
\begin{equation}\label{g}
g(r)=\big(1-\frac{2M}{r}+\frac{cm_p^2}{r^2}+\frac{c^\prime m_p^2M}{r^3}+\frac{c^{\prime\prime}m_p^2M^2}{r^4}\big)^{-1},
\end{equation}
where
\begin{equation}
\begin{split}
c=\frac{D\sum A_s}{960\pi}&=0.004,\quad c^\prime=\frac{\sum A_s(B_s-1)}{320\pi}=0.029,\\
&c^{\prime\prime}=\frac{32\sum q_s-5\sum A_sB_s}{1440\pi}=1.358.
\end{split}
\end{equation}
Subtracting \eqref{first} from \eqref{second} leads to
\begin{equation}
\frac{(fg)^\prime}{rfg^2}=0\quad\Rightarrow\quad fg=constant,
\end{equation}
which results in $f=g^{-1}$. These obtained forms of $f$ and $g$ also satisfiy the relation of \eqref{third}. Thus, the quantum-corrected Schwarschild metric becomes
\begin{equation}\label{mmetric}
ds^2=-f(r)dt^2+f^{-1}(r)dr^2+r^2(d\theta^2+\sin^2\theta d\phi^2),
\end{equation}
with
\begin{equation}\label{function}
f(r)=\big(1-\frac{2M}{r}+\frac{cm_p^2}{r^2}+\frac{c^\prime m_p^2M}{r^3}+\frac{c^{\prime\prime}m_p^2M^2}{r^4}\big).
\end{equation}

The first and second terms of corrections in $f$ is due to the vacuum polarization alone, while the last term is induced by a combined effect of the vacuum polarization and the conformal anomaly. Examining the roots of $f$ indicates that this function has two positive real roots, which represents the existence of two horizons. The bigger root corresponding the outer horizon is
\begin{equation}
r_+\simeq\frac{2M}{1+a\big(\dfrac{m_p}{M}\big)^2}, \quad a=\frac{c}{4}+\frac{c^\prime}{8}+\frac{c^{\prime\prime}}{16}.
\end{equation}
The outer horizon is located near the classical Schwarzschild horizon. The other root representing the inner horizon is
\begin{equation}
r_-\simeq\frac{\big(\dfrac{c^{\prime\prime}}{2}\dfrac{M}{m_p}\big)^{\frac{1}{3}}}{1-b\big(\dfrac{c^{\prime\prime}}{2}\dfrac{M}{m_p}\big)^{\frac{1}{3}}\dfrac{m_p}{M}}l_p,\quad b=\frac{2c^\prime+c^{\prime\prime}}{3c^{\prime\prime}}.
\end{equation}
If the gravitational mass $M$ is of the order of  an ordinary black hole, then the inner horizon will be at a distance much larger than the Planck length. Therefore, quantum effects cause the structure of the Schwarzschild geometry to alter so that there are two event horizons instead of one. In the following, we will examine how these changes affect the gravitational collapse and the formation of black holes.
\section{The semi-classical collapse of a thin shell of dust}\label{section 3}
In this section, the gravitational collapse of a spherical shell is investigated in the presence of quantum vacuum effects. The space-time outside the shell is described by the modified Schwarzschild geometry of \eqref{mmetric}. Space-time inside the shell is flat, so the metric in this region will have the Minkowski form. Due to the absence of space-time curvature in the inner region of the shell, there is no contribution from the conformal anomaly and the vacuum polarization. The only non-zero quantum effect is the Casimir effect, which is negligible compared to the two considered effects, so we withdraw it from the calculations.

In order to obtain the motion equations of the shell, the matching or junction conditions for geometry of the interior and exterior regions must be examined \cite{Israel, Barrabes, Poisson}. If a hypersurface $\Sigma$ divides the space-time into regions of $V^+$ and $V^-$ with coordinates $x^\mu_+$ and $x^\mu_-$ and metrics $g^+_{\mu\nu}$ and $g^-_{\mu\nu}$, respectively, then the metric induced on $\Sigma$ by each of these two regions is
\begin{equation}
ds^2_\Sigma=h^\pm_{ab}dy^a_\pm dy^b_\pm,\quad h^\pm_{ab}=g^\pm_{\mu\nu}e^\mu_{\pm a} e^\nu_{\pm b},\quad a,b=1, 2, 3
\end{equation}
where $y^a_\mp$ are the parameters characterizing $\Sigma$ and $e^\mu_{\pm a}=\frac{\partial x^\mu_\pm}{\partial y^a_\pm}$ are the tangent vectors to the hypersurface. If $[h_{ab}]=h^+_{ab}-h^-_{ab}=0$, then the first junction condition is satisfied. In this way, the metric changes continuously in passing through each of the two regions. If the other condition also holds, then there is a smooth  joining of $g^\pm_{\mu\nu}$ at $\Sigma$.

The second condition is related to the changes of the extrinsic curvature tensor. If $n^\pm_\mu$ is the normal vector on $\Sigma$, then the extrinsic curvature tensor on each side is defined as
\begin{equation}
K^\pm_{ab}=\nabla_\nu n^\pm_\mu e^\mu_{\pm a} e^\nu_{\pm b}.
\end{equation}
The difference of the extrinsic tensor of the hypersurface in two sides determines whether the union of the regions of $V^+$ and $V^-$ form a single space-time or not (the second junction condition). If this difference is equal to zero, in that $[K_{ab}]=0$, then the answer to the question is positive. Otherwise, there is a non-zero stress-energy tensor on $\Sigma$ with the form
\begin{equation}\label{st tensor1}
S_{ab}=[K]h_{ab}-[K_{ab}],\quad K= K_{ab}h^{ab}.
\end{equation}

Now, we study the problem of gravitational collapse of a shell. In this regard, regions of $V^+$ and $V^-$ correspond to the modified Schwarzschild and Minkowski space-times, respectively, and $\Sigma$ represents the spherical shell. The metric induced on the shell by each of the two inner and outer areas has the following single form
\begin{equation}
ds^2_\Sigma=-d\tau^2+R^2(\tau)(d\theta^2+\sin^2\theta d\phi^2),
\end{equation}
where $\tau$ is the time in the co-moving frame and $R(\tau)$ is the radius of the shell as a function of time. Having the same metric form on both sides implies that $[h_{ab}]=0$, which indicates that the first junction condition holds. Therefore, in passing from $V^+$ to $V^-$ and vice versa, the metrics are joined  continuously.

Considering that regions of  $V^+$ and $V^-$ are separated from each other by a shell of matter, it is expected that the second junction condition is not satisfied, in that $[K_{ab}]\ne0$. Indeed, the distribution of matter for a spherical shell of pressureless dust is described by the stress-energy tensor
\begin{equation}\label{st tensor2}
S_{ab}=\sigma u_au_b,
\end{equation}
in which $\sigma$ is the surface density and $u_a$ is the four-velocity of the shell.

Calculating the extrinsic curvature tensor on both sides of the shell leads to
\begin{equation}\label{ex tensor}
\begin{split}
&K^\tau_{\pm\tau}=\frac{\dot{F}_\pm}{\dot{R}},\quad K^\theta_{\pm\theta}=K^\phi_{\pm\phi}=\frac{F_\pm}{R},\\
&F_+=\sqrt{\dot{R}^2+f(R)},\quad F_-=\sqrt{\dot{R}^2+1},
\end{split}
\end{equation}
where the dot denotes the derivative with respect to $\tau$. Now, using relations of \eqref{ex tensor} and comparing \eqref{st tensor1} and \eqref{st tensor2} results in
\begin{equation}
S^\tau_\tau=\frac{F_+-F_-}{4\pi R}=-\sigma,\quad S^\theta_\theta=S^\phi_\phi=\frac{F_+-F_-}{8\pi R}+\frac{\dot{F}_+-\dot{F}_-}{8\pi\dot{R}}=0.
\end{equation}
Integrating the relationship for $S^\theta_\theta$ gives $(F_+-F_-)R=constant$, and substituting this into the equation for $S^\tau_\tau$ yields
\begin{equation}\label{fsheq}
4\pi R^2\sigma\equiv m=-constant,
\end{equation}
which reveals the connection between the obtaind constant and the rest mass of the shell, i.e., $m$.
Therefore, we get
\begin{equation}\label{ssheq}
\sqrt{\dot{R}^2+1}-\sqrt{\dot{R}^2+f(R)}=\frac{m}{R}.
\end{equation}
The relations of \eqref{fsheq} and \eqref{ssheq} are the equations of the motion, which govern on the dynamics and time evolution of the shell. Solving the differential equation of \eqref{ssheq} is impossible in general, of course putting $\dot{R}=0$ and finding turning points provide useful information about the final state of a collapsing shell ($\dot{R}<0$).

\begin{figure}[t!]
\centering
\includegraphics[width=12cm]{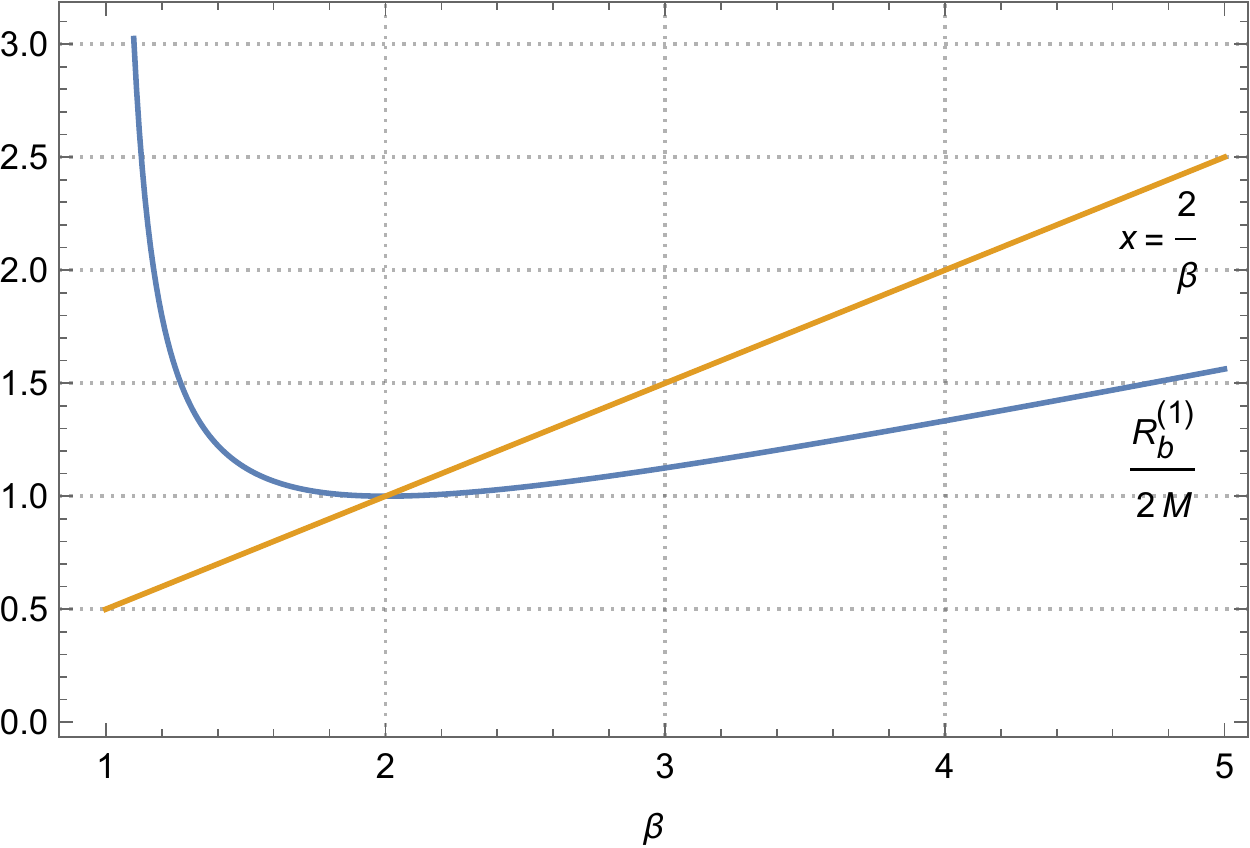}
\caption{The blue curve is a plot of $R^{(1)}_b$ in terms of $\beta$ for a shell with $\alpha=10^{-38}$. The orange line represents the condition of \eqref{second condition}, which is equivalent to $x=2/\beta$. Only those points of the blue curve are acceptable as soultios to the equation of \eqref{turning point eq}, which are located on the left side of the orange line.}
\label{first br}
 \end{figure}

Letting $\dot{R}=0$, the equation of \eqref{ssheq} takes the form
\begin{equation}\label{turning point eq}
1-\frac{m}{R}=\sqrt{1-\frac{2M}{R}+\frac{cm_p^2}{R^2}+\frac{c^\prime m_p^2M}{R^3}+\frac{c^{\prime\prime}m_p^2M^2}{R^4}}.
\end{equation}
The roots of this equation, if any, must apply in the following conditions
\begin{subequations}
\begin{equation}\label{first condition}
R\geq r_+\quad or\quad  R\leq r_-,
\end{equation}
\begin{equation}\label{second condition}
R\geq m.
\end{equation}
\end{subequations}
By putting $x=2M/R$, $\alpha=m_p/M$, $\beta=m/M$ and squaring the both sides of \eqref{turning point eq}, one finds the cubic equation
\begin{equation}\label{cubic equation}
\frac{c^{\prime\prime}}{16}\alpha^2x^3+\frac{c^{\prime}}{8}\alpha^2x^2+(\frac{c}{4}\alpha^2-\frac{\beta^2}{4})x+\beta-1=0.
\end{equation}
This equation has one or two positive real roots for different values of $\beta$. For $\beta>1$, the greater root indicates the existance of a turning point equal to
\begin{equation}\label{first bounce radious}
R^{(1)}_b\simeq\frac{\beta^2-\alpha c}{4(\beta-1)}M.
\end{equation}
The graph of $R^{(1)}_b$ in terms of $\beta$ for a shell with the sun's mass ($\alpha=10^{-38}$) is shown in figure \ref{first br}. As it can seen, $R^{(1)}_b$ is always greater than $r_+$, so it affirms the condition of \eqref{first condition}. The condition of \eqref{second condition} in terms of $x$ is equivalent to $x=2/\beta$, which is shown with the orange line. This requires that the value of $\beta$ must be smaller than $\beta\simeq2-\alpha c/2$ for a turning point to exist.
In this case, the gravitational collapse will stop before any horizon or singularity is formed. It should be noted that this happens when the initial radius of the shell is larger than the value of the turning point, i.e., ($R_0>R^{(1)}_b$). Indeed, according to figure \ref{first br} and also equation of \eqref{first bounce radious}, for values of $\beta$ that are close to $1$, the value of $R^{(1)}_b$ will be very large and certainly the initial radius of such a shell will be smaller than that. But for those  values of $\beta$ that are close to $2$, the turning point is located at a close distance from the outer horizon (slightly greater), and therefore shells with these values of $\beta$ will never turn into black holes, since the creation of the outer horizon is canceled.

\begin{figure}[t!]
\centering
 \includegraphics[width=12cm]{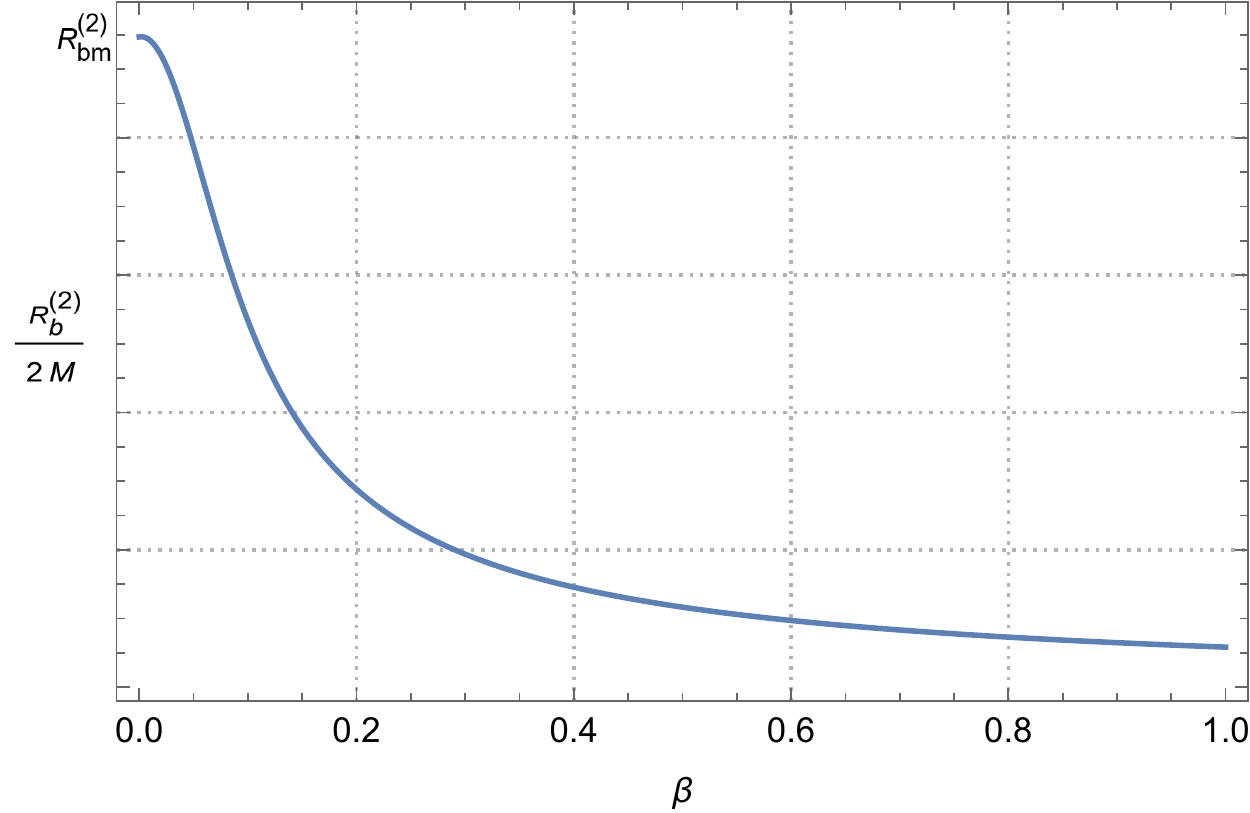}
 \caption{The plot of  $R^{(2)}_b$ in terms of $\beta$. For all values of $\beta$, $R^{(2)}_{b}$ is smaller than the inner horizon. If the gravitational mass is in the order of astrophysical stars, $R^{(2)}_{b}$ will be much larger than the Planck length. The maximum value of $R^{(2)}_{bm}$ takes place when the rest mass of the shell is equal to zero.}
\label{second br}
\end{figure}

\begin{figure}[t!]
\centering
 \includegraphics[width=12cm]{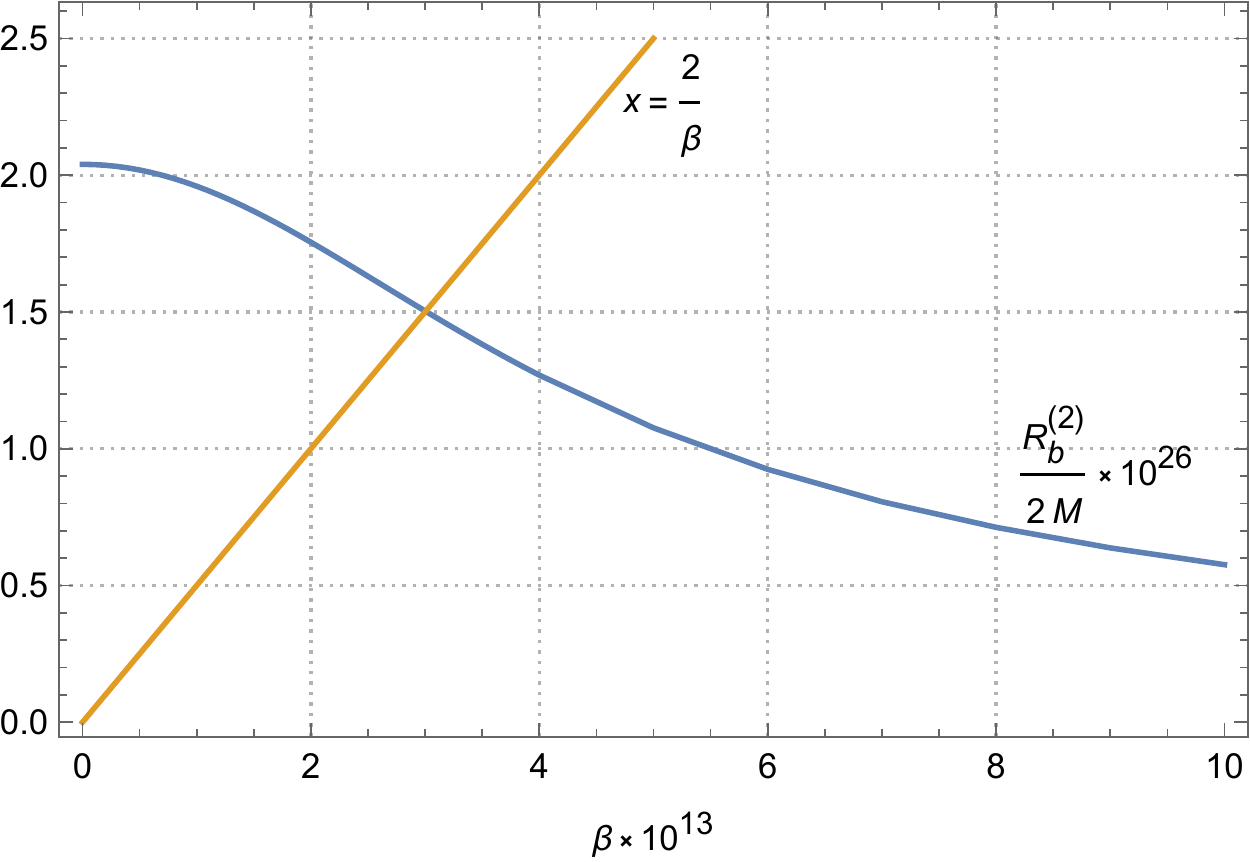}
 \caption{The condition of \eqref{second condition} represented through the orange line for a shell of $\alpha=10^{-38}$. It implies that there is a turning point only for $\beta\le3\times10^{-13}$ . For such a shell with this restriction, the final outcome is creation of a non-singular black hole. This limitation cause the gravitational collapse to finish with a sigular state in most cases.} 
\label{second br2}
\end{figure}

The other real and positive root of \eqref{cubic equation}, which exists for all values of $\beta$, is located at a smaller radius than $r_-$, and therefore the condition of \eqref{first condition} holds. In figure \ref{second br}, a schematic plot of the second turning piont, $R^{(2)}_b$, is shown in terms of $\beta$. $R^{(2)}_b$ decreases with the increase of $\beta$, so for $\beta = 0$, that is, when the rest mass of the shell vanishes, $R^{(2)}_b$ takes its largest value, i.e., $R^{(2)}_{mb}$. This value occurs for $f(R)=1$, which is approximately equal to
\begin{equation}
R^{(2)}_{mb}\simeq\bigg[\big(\frac{c^{\prime\prime}}{16}\alpha^2\big)^{-\frac{1}{3}}-\frac{2c^\prime}{3c^{\prime\prime}}\bigg]l_p,
\end{equation}
which will be much larger than the Planck radius for shells with an enormous value of gravitational mass.
Examining the condition of \eqref{second condition} demonstrates that very few points from the graph in figure \ref{second br} satisfy this constraint. In fact, for a shell with a gravitational mass equal to the mass of the sun, $R^{(2)}_{b}$ only exists for $\beta\le3\times10^{-13}$ (see figure \ref{second br2}). Therefore, only for shells with very small rest mass and close to zero, there will be the turning point of $R^{(2)}_{b}$. In this way, the outer and inner horizons are formed, but the collapse stops before reaching the final singularity, leading to the creation of a non-singular black hole. After this event, the radius of the shell begins to increase, and a bounce takes place. By rewriting the metric of \eqref{mmetric} in non-singular coordinates at $r_-$, such as Eddington-Finkelstein or Painleve-Gullstrand, and studying the geodesic equation of the shell, it can be found that from the point of view of an external observer, it will take a long time for the radius of the shell to back to the position of the inner horizon. Thus from the outside, it seems like a stable black hole \cite{f4}.

But in most cases, i.e., for $\beta>3\times10^{-13}$, there is no bouncing point and the final fate of the gravitational collapse will be a singularity. This means that for a gravitational collapse with the mass of an ordinary star, the prevention from the formation of the singularity is impossible, even in the presence of the quantum vacuum effects. Of course, quantum effects also cause some modifications, including the change in the geometry of Schwarzschild space-time, which leads to two horizons for black holes. However, for genuine black holes, detecting the effects of quantum corrections on the geometry through astrophysical observations is challenging since quantum corrections insignificantly affect the geodesics and orbits, such as the photon sphere. Whether these changes are observable requires more research.

It should be noted that all calculations in this section are considered for shells with a gravitational mass equivalent to ordinary stars ($\alpha=10^{-38}$). Since for light and sufficiently small shells, it is no longer possible to ignore the non-diagonal components of the stress-energy tensor of quantum fields, i.e., Hawking radiation. The temperature of the Hawking radiation of a black hole has an inverse relationship with its mass, in other words, the smaller the black hole's mass, the higher its Hawking radiation will be. Therefore, for small black holes \cite{Shafiee}, evaporation is to the extent that their gravitational mass should be considered as a function of time, while it takes $10^{67}$ years to completely evaporate a black hole with the mass of the sun. In this regard, the considered stress-energy tensor, and consequently  Einstein's equations of \eqref{e equation} all need to be altered and rewritten, because all subsequent calculations are no longer valid. Of course, considering the Hawking radiation for the proposed model (a shell with the sun's mass), there is a very small shift in the values of $c$, $c^\prime$, and $c^{\prime\prime}$, which will not affect the obtained results.

Due to the negative energy of the ongoing part of the Hawking radiation, the outer horizon begins to shrink. While this part converts to positive energy modes after crossing the inner horizon. This results in the growth of the inner horizon, so the trapped surface between the horizons get small and smaller until it disappears by coinciding and uniting the two horizons with each other \cite{e1, f4}. This is not going to happen from a Planck size region, but from a macroscopic region, which can contain the residual information that did not escape with the Hawking radiation. Thus in this scenario, it seems that  the information is not lost, and they are preserved through a large remnant after the final stages of evaporation.

It is worth mentioning that in general, to describe precisely the final stages of black hole evaporation, one needs a non-perturbative theory of quantum gravitation. With the progress of the gravitational collapse, the quantum effects appear stronger, leading to modifications in the form of equations. Hence, we must be careful about the validity of the semi-classical approach when we talk about the final state of the gravitational collapse and the evaporation process.
\section{Conclusion and remarks}
This prediction of the quantum theory that the energy of the vacuum state has a non-vanishing value is a motivation to revise the solutions of Einstein's equations, especially black holes. In this regard, we analyzed the problem of gravitational collapse by considering the backreaction of quantum fields. Quantum effects create modifications in the structure of the Schwarzschild geometry that can affect gravitational collapse. These changes predict the existence of two event horizons for black holes. An outer horizon, which is the same as the horizon of classical black holes with some modifications, and the other is an inner horizon whose existence is entirey derived through quantum effects. Investigating the collapse of a spherical shell by considering this modified geometry showed that under certain conditions, it is possible to stop the collapse before the formation of a singularity or even a horizon.

These conditions are restricted when the ratio of the rest mass of the shell to its gravitational mass, i.e., $\beta$, takes specified values. In one case, the collapse ceases, even before the formation of any horizon, and therefore there is no longer any possibility of a black hole creation. In the second case, the outer and inner horizons form, but the collapse stops before the singularity creates, and then the shell begins an expansion. To an outside observer, it takes a very long time for the radius of the shell to return to the location of the inner horizon, so this non-singular black hole looks like a stable black hole from the outside. In the real world, there is no chance for the stated conditions to be realized, so in the semi-classical case, as in the classical picture, despite the differences that exist, the collapse of a sufficiently heavy object leads to the formation of a gravitational singularity.

The equation of \eqref{cubic equation} was examined in the case where the value of $\alpha$ is so small, i,e. for the the mass of the sun. If $\alpha$ does not satisfy this condition, then the obtained results will no longer be valid. In this regard, the Hawking radiation cannot be neglected anymore, and the stress-energy tensor has another form, which must be written in its correct appearance. Even if the mass of the shell is smaller than a definite limit \cite{Shafiee}, then the Hawking radiation is so large and intense that the mass of the shell will be a function of time, which will lead to fundamental changes in the form of the equations and the results. The evaporation process leads to the disappearing of the trapped surface through the unification of the outer and inner horizons, resulting in an extremal black hole \cite{final1, final2}. 

 Also, as can be seen from \eqref{function}, the corrections induced by the conformal anomaly and the vacuum polarization alter the value of the surface gravity and consequently the temperature of Hawking radiation. In other words, quantum effects impress each other through the modifications they create. On the other hand, with the time progression and the further reduction of mass by Hawking radiation, according to the equations of \eqref{ca} and \eqref{vp}, other quantum effects are also modified. What is evident in this regard is that this action and reaction has an infinite sequence, and probably the problem cannot be examined perturbatively and to clarify the issue, a quantum gravitational approach is required. This subject is under investigation and will be addressed more in the future research.

To make the scenario more realistic, the gravitationaal collapse can be considered for a sphere containing matter instead of a spherical shell. In \cite{Arfaei}, this issue has been studied for a ball of dust in the presence of the conformal anomaly, and it has been shown that the formation of a non-singular black hole is possible in some conditions. Also, the subject can be generalized to other types of black holes, especially rotating ones.  Since genuine black holes have the property of rotation, investigating the subject in this situation can help to identify the internal structure of black holes through the data obtained from gravitational waves by LIGO \cite{Zeeya}. In this case, it is possible to compare the data with various models and choose the model that is more consistent with the data. It is clear that there is a long way ahead in this regard, and for a final picture to emerge, there is no other way than data gathering and further research.

\bibliographystyle{JHEP}
\bibliography{References}

\end{document}